\documentclass[runningheads,a4paper]{llncs}

\usepackage[english]{babel}
\usepackage[utf8x]{inputenc}
\usepackage[T1]{fontenc}

\usepackage{amsmath}
\usepackage{graphicx}
\usepackage[colorinlistoftodos]{todonotes}
\usepackage[colorlinks=true, allcolors=blue]{hyperref}

\authorrunning{Dragos Strugar et al.}

\title{Reputation in M2M Economy}
\subtitle{A Case Study of Electric Autonomous Vehicles}

\author{
Dragos Strugar, Rasheed Hussain, JooYoung Lee, Manuel Mazzara, Victor Rivera\\
\institute{Innopolis University, Russian Federation}
\{d.strugar, r.hussain, j.lee, m.mazzara, v.rivera\}@innopolis.ru
}

\begin{document}
\maketitle
\section{Electric Autonomous Vehicles and M2M Economy}
Triggered by modern technologies, our possibilities may now expand beyond the unthinkable. Cars externally may look similar to decades ago, but a dramatic revolution happened inside the cabin as a result of their computation, communications, and storage capabilities. With the advent of Electric Autonomous Vehicles (EAVs) \cite{presentation2015future}, Artificial Intelligence and ecological technologies found the best synergy. Several transportation problems may be solved (accidents, emissions and congestion among others), and the foundation of Machine-to-Machine (M2M) economy could be established, in addition to value-added services such as infotainment (information and entertainment).

\section{The Issue of Reputation}
In the world where intelligent technologies are pervading everyday life, software and algorithms play a major role. Software has been lately introduced in virtually every technological product available on the market, from phones to television sets to cars and even housing. Artificial Intelligence is one of the consequences of this pervasive presence of algorithms. The role of software is becoming dominant and technology is, at times pervasive, of our existence. Concerns, such as privacy and security, demand high attention and have been already explored to some level of detail. However, intelligent agents and actors are often considered as perfect entities that will overcome human error-prone nature. This may not always be the case and we advocate that the notion of \textit{reputation} is also applicable to intelligent artificial agents, in particular to EAVs.

\section{M2M and Reputation}
EAVs, like every intelligent actor, are affected by the quality and quantity of the actions they perform. With the current level of technology (and possibly for a long time to come) there is no reason to assume that EAVs should behave perfectly without any abnormal deviation from the expected behavior. Also, as learning agents, their development is affected by the environment and the training context, meaning some actors can act more reliably than others. This aspect is even more relevant when EAVs interact with humans to take decisions, for example, when the driving is supervised by humans who can intervene at any time. We advocate for the need to define \textit{Reputation} for EAVs. Mathematical models of reputation do exist and can be applied here~\cite{Lee:2013:MRP:2492517.2492663}~\cite{journals/jucs/LeeDODBWJ12}. In particular, we plan to adapt our model, as defined in~\cite{Melnikov18}, to a specific case study.

\section{Case study}

The case study described in~\cite{Strugar18} consists of  EAVs which are able to operate based on M2M charging (wired or wireless) and billing without human mediation. The billing framework is based on IOTA \cite{medium2017introductiontoiota}, the cryptocurrency which aims to become the backbone of the Internet-of-Things (IoT) industry by providing feeless transactions at scale. This approach demands the market to offer appropriate charging services without involving humans. To date, several mechanisms have been proposed by researchers to deal with the charging and auditability \cite{Zhao2015,Au2014} \cite{Hussain2017,Rezaeifar2017}; however, the state-of-the-art mechanisms of charging and billing do not meet the M2M economy requirements as they often impose service fees for value transactions that may also endanger users and their location privacy. 

In~\cite{Strugar18}, Strugar et al. aimed at filling this gap and envisioned a new charging architecture and a billing framework for EAVs to enable M2M transactions via the use of Distributed Ledger Technology (DLT). Our objective in this paper is, instead, to further discuss the aspects of reputation. Self-driving cars will interact with each other, external entities, and with charging stations. This communication is based on mutual trust among the communicating entities and therefore requires reputation management. EAVs will react to approaching entities based on the reputation model and take actions accordingly.

\begin{figure}
\includegraphics[scale=0.4]{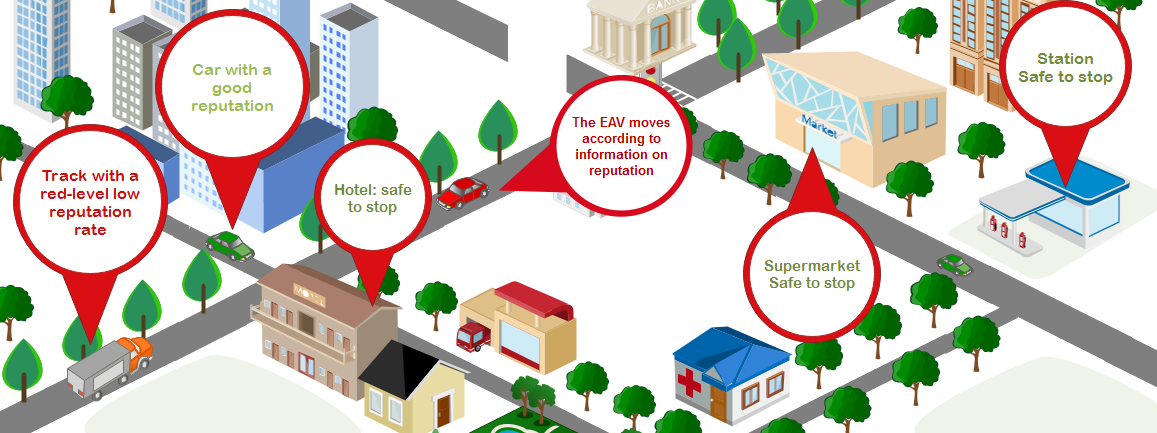}
\centering
\caption{Reputation Management}
\label{fig:scenario}
\end{figure}

Figure \ref{fig:scenario} depicts a possible reputation management scenario. Entities such as EAVs, charging stations and other facilities have reputation scores which are used as a measure of trust. These scores affect the actions and decisions of other entities in their surrounding. Consequently, entities with the higher score tend to have more interactions with others.

\section{Reputation Model}
Trust can be seen as the quality and quantity of interactions among actors: the more interactions occur between two parts, the more one trusts the other. In \cite{Melnikov18}, the Dynamic Interaction Based Reputation Model (DIB-RM) was introduced to capture this dynamic property of trust. 

DIB-RM is an interaction-based model among users of a community over time that seems to be applicable to any kind of intelligent actors, be it humans or artificial agents, including EAVs. The model computes a reputation value for each actor on the system combining different dynamic factors. The reputation value is updated at each interaction, and it has been shown effective in dynamic environments: the model can update the reputation value of users while they perform other actions. 

The model will be adapted to the case study and validated through simulations, at least until the full deployment of the technology is realized.

\bibliographystyle{alpha}
\bibliography{bibliography}

\newcommand{\etalchar}[1]{$^{#1}$}
\begin{thebibliography}{RHKO17}

\bibitem[ALF{\etalchar{+}}14]{Au2014}
M.~H. Au, J.~K. Liu, J.~Fang, Z.~L. Jiang, W.~Susilo, and J.~Zhou.
\newblock A new payment system for enhancing location privacy of electric
  vehicles.
\newblock {\em IEEE Transactions on Vehicular Technology}, 63(1):3--18, Jan
  2014.

\bibitem[AM18]{Melnikov18}
Victor Rivera JooYoung Lee Luca~Longo Almaz~Melnikov, Manuel~Mazzara.
\newblock Towards dynamic interaction-based reputation models.
\newblock In {\em Proceedings of the 32nd International Conference on Advanced
  Information Networking and Applications Workshops (WAINA)}, 2018.

\bibitem[AS15]{presentation2015future}
Muhammad Azmat and Clemens Schuhmayer.
\newblock Future scenario: Self driving cars - the future has already begun.
\newblock
  \url{https://www.researchgate.net/publication/278329250_Future_Scenario_Self_Driving_Cars_-_The_future_has_already_begun},
  2015.

\bibitem[DS18]{Strugar18}
Manuel Mazzara Victor~Rivera Dragos~Strugar, Rasheed~Hussain.
\newblock M2m billing for electric autonomous vehicles.
\newblock \url{https://arxiv.org/abs/1804.00658}, 2018.

\bibitem[HSK{\etalchar{+}}17]{Hussain2017}
Rasheed Hussain, Junggab Son, Donghyun Kim, Michele Nogueira, Heekuck Oh,
  Alade~O. Tokuta, and Jungtaek Seo.
\newblock Pbf: A new privacy-aware billing framework for online electric
  vehicles with bidirectional auditability.
\newblock {\em Wireless Communications and Mobile Computing}, 2017:17, 2017.

\bibitem[LDO{\etalchar{+}}12]{journals/jucs/LeeDODBWJ12}
JooYoung Lee, Yue Duan, Jae~C. Oh, Wenliang Du, Howard Blair, Lusha Wang, and
  Xing Jin.
\newblock Social network based reputation computation and document
  classification.
\newblock {\em J. UCS}, 18(4):532--553, 2012.

\bibitem[LO13]{Lee:2013:MRP:2492517.2492663}
Joo~Young Lee and Jae~C. Oh.
\newblock A model for recursive propagations of reputations in social networks.
\newblock In {\em Proceedings of the 2013 IEEE/ACM International Conference on
  Advances in Social Networks Analysis and Mining}, ASONAM '13, pages 666--670,
  New York, NY, USA, 2013. ACM.

\bibitem[RHKO17]{Rezaeifar2017}
Zeinab Rezaeifar, Rasheed Hussain, Sangjin Kim, and Heekuck Oh.
\newblock A new privacy aware payment scheme for wireless charging of electric
  vehicles.
\newblock {\em Wireless Personal Communications}, 92(3):1011--1028, Feb 2017.

\bibitem[Sch17]{medium2017introductiontoiota}
Dominic Schiener.
\newblock Introduction to iota cryptocurrency.
\newblock
  \url{https://blog.iota.org/a-primer-on-iota-with-presentation-e0a6eb2cc621},
  2017.

\bibitem[ZZWZ15]{Zhao2015}
T.~Zhao, C.~Zhang, L.~Wei, and Y.~Zhang.
\newblock A secure and privacy-preserving payment system for electric vehicles.
\newblock In {\em 2015 IEEE International Conference on Communications (ICC)},
  pages 7280--7285, June 2015.

\end{thebibliography}

\end{document}